\documentclass[12pt,preprint]{aastex}

\newcommand\simgt{\lower.5ex\hbox{$\; \buildrel > \over \sim \;$}}
\newcommand\simlt{\lower.5ex\hbox{$\; \buildrel < \over \sim \;$}}

\newcommand\IM{\langle I/{\cal M}\rangle}
\newcommand\edot{\langle \dot{e}\rangle}
\newcommand\edotx{\langle \dot{e}(x)\rangle}
\newcommand\edotr{\langle \dot{e}(r)\rangle}

\newcommand\FDF{{\bf F}_{\rm DF}}
\newcommand\calL{{\cal L}}
\newcommand\Msun{M_\odot}
\newcommand\vturb{v_{\rm turb}}
\newcommand\TkeV{T_{\rm keV}}
\newcommand\Tiso{T_{\rm iso}}
\newcommand\Tsonic{T_{\rm sonic}}
\newcommand\kms{{\rm\;km\;s^{-1}}}
\newcommand\pcc{{\rm\;cm^{-3}}}
\newcommand\ergcms{{\rm\;erg\;cm^{-3}\;s^{-1}}}
\newcommand\rmin{r_{\rm min}}

\newcommand\sles{\lower2pt\hbox{$\buildrel {\scriptstyle <}
   \over {\scriptstyle\sim}$}}
\newcommand\sgreat{\lower2pt\hbox{$\buildrel {\scriptstyle >}
   \over {\scriptstyle\sim}$}}
\shortauthors{Kim, El-Zant, \& Kamionkowski}
\shorttitle{Dynamical Friction and Cooling Flows}

\begin{document}

\title{Dynamical Friction and Cooling Flows in Galaxy Clusters}

\author{Woong-Tae Kim}
\affil{Astronomy Program, SEES, Seoul National University, 
Seoul 151-742, Korea}
\affil{Harvard-Smithsonian Center for Astrophysics, 60 Garden Street,
Cambridge, MA 02138, USA}
\email{wkim@astro.snu.ac.kr}

\author{Amr A. El-Zant}
\affil{CITA, University of Toronto, Ontario M5S 3H8, Canada}

\and

\author{Marc Kamionkowski}
\affil{Mail Code 130-33, California Institute of Technology,
Pasadena, CA 91125, USA}

\begin{abstract}
We investigate a model of galaxy clusters in which the hot intracluster
gas is efficiently heated by dynamical friction (DF) of galaxies.  
We allow for both subsonic and supersonic motions of galaxies and use the 
gravitational drag formula in a gaseous medium presented by Ostriker (1999).
The energy lost by the galaxies is either redistributed locally or into a 
Gaussian centered on the galaxy.
We find that the condition of hydrostatic equilibrium and 
strict energy balance yields a trivial isothermal solution $\Tiso$, 
independent of radius, or rising temperature distributions provided
$\Tiso/\gamma<T<\Tiso$, where $\gamma$ is the adiabatic index of the gas. 
The isothermal temperature corresponds  to the usual scaling relation
between the gas temperatures and the velocity dispersions of galaxies.
However the minimal temperature associated with the rising  solutions 
is $\sim \frac{1}{2} T_{\rm vir}$, larger than that inferred from 
observations,  the  radial distribution of galaxy masses notwithstanding.
Heating by supersonically moving galaxies cannot suppress 
thermal instability, although it can lengthen the growth time
up to the level comparable to the ages of clusters when Mach number
of galaxies is less than about two.
We show using numerical hydrodynamic simulations that DF-induced heating
is generally unable to produce stable equilibrium cores by evolving 
arbitrary non-equilibrium clusters, although it can lengthen the cooling time.
We conclude that DF-induced heating {\it alone} is an unlikely solution 
to the cooling flow problem, although it can still be 
an important heat supplier,  considerably delaying cooling catastrophe. 
We discuss other potential consequences of DF of galaxies in 
galaxy clusters.
\end{abstract}


\section{Introduction}

Galaxy clusters are the largest gravitationally bound structures in 
the universe.  They typically contain several hundreds to thousands
of galaxies orbiting in a gravitational potential well formed primarily by
dark matter (e.g., \citealt{bah77}).  
They are also filled with hot gas with $T\sim2-10$ keV that loses thermal
energy prolifically by emitting X-rays.
In the absence of any heat sources, the radiative cooling in the cores 
of rich clusters would result in a cooling flow in
which gas settles in the gravitational potential and drops out as cold
condensations (e.g., \citealt{fab94}).
However, recent high resolution {\it Chandra} and {\it XMM-Newton}
observations have not shown the expected signature of gas cooling below
one-third of the virial temperature 
\citep{bla01,pet01,pet03,tam01,joh02,all01}.
This strongly suggests that there must be some source of heat that 
manages to balance the radiative cooling and thus prevents the mass dropout
at the central regions of galaxy clusters. 
Most theoretical models proposed so far for cluster heating 
fall into the following two groups:
(1) energy injection via radiations, jets, outflows, bubbles, or
sound waves from a central active galactic nucleus 
(\citealt{cio01,chu02,kai03,rus04,roy04,vec04} and references therein); 
(2) diffusive transport of heat from the outer regions of the cluster to the 
center via conduction \citep{tuc83,bre88,nar01,voi02,zak03,kim03a} or
turbulent mixing \citep{cho03,kim03b,voi04,den05}.

Although less well studied than those mentioned above, 
there are clearly other energy sources 
in the cluster environments. 
These include kinetic energy in the orbital motions of galaxies
\citep{mil86,jus90},
gravitational potential energy of the gas \citep{mar01,fab03},
feedback from intracluster supernovae \citep{dom04},
dissipation of turbulent energy driven by infall of
subclusters \citep{fuj04}, etc.
Amount of available energy in each source is comparable to or larger than
thermal energy of the intracluster medium (ICM), so that 
radiative cooling would be easily offset if there exist physical 
mechanisms that can tap the energy from the sources
and convert it into thermal energy in the ICM.

One such process is dynamical fiction (DF) which occurs when a galaxy 
moving around a cluster center 
interacts with its gravitationally induced wake 
in the ICM (e.g., \citealt{dok64,hun71,rud71,rep80}).
Because of the gravitational drag, the galaxy loses some of its kinetic 
energy.  Turbulence in the ICM generated by the superposition of the wakes
of many galaxies or by other processes will absorb the lost kinetic energy
and, in the presence of tangled magnetic fields and/or viscosity,
turns it into heat at the dissipation scales (e.g., \citealt{dei96}). 
\citet{mil86} and \citet{jus90} independently estimated the heating rates 
by DF of galaxies and found that the DF-induced heating can fairly well 
balance radiative cooling of a rich cluster, provided the mass-to-light 
ratio is about 20. 
Using updated data for the properties of galaxies and gas
in the Perseus cluster and employing Monte-Carlo approaches,
\citet[hereafter EKK04]{elz04} very recently
showed that the total power generated by DF within the cooling 
radius is comparable to radiative loss from that region if
the average mass-to-light ratio of galaxies is about 10.
They also noted that this mechanism is self-regulating in the sense that
gas would not be overheated since the DF of galaxies becomes ineffective 
when it attains high enough temperature for which galaxy motions become
subsonic (see also \citealt{mil86}). 

While the results of the aforementioned work suggest that 
the {\it total} power supplied by the motions of member galaxies is 
comparable to X-ray luminosity of a typical rich cluster,
it is sill uncertain whether it can balance
the radiative cooling {\it locally} as well.
If the DF of galaxy motions is to serve as a primary heat supplier 
in clusters in equilibrium, 
observed density and temperature profiles of clusters should be
a natural consequence of the local heat 
balance between radiative cooling and DF-induced heating.
Since optically-thin, X-ray emitting gas is prone to thermal
instability (e.g., \citealt{fab94}), it is also an interesting question
whether the DF-induced heating is able to suppress
thermal instability completely or at least lengthen its growth time
to the level comparable to the ages of clusters;
even an equilibrium cluster would otherwise be subject to
a mass dropout at its center \citep{kim03a}.

In this paper, we take one step further from \citet{mil86} and EKK04
to investigate equilibrium cluster models in
which DF-induced heating is a main heating mechanism.
We will first construct density and temperature distributions of the
gas in clusters that are in strict hydrostatic and thermal equilibrium
and compare them with observed profiles.
We note that recent X-ray data from {\it BeppoSAX}, {\it Chandra}, and 
{\it XMM-Newton}
observations indicate that gas temperature in rich clusters rapidly increases 
with radius for $r\simlt (0.05-0.08)r_{180}$, remains roughly isothermal up to 
$r\sim 0.2r_{180}$, and then gradually declines farther out
(\citealt{ett00,deg02,pif05,vik05}, see also \citealt{mar98} for {\it ASCA}
results).
Here, $r_{180}$ is a virial radius the interior of which has 
the mean density of gas equal to 180 times the critical density.
While declining temperature profiles at large radii may provide
important clues to cosmological structure formation, they appear
not to be directly relevant to the cooling flow problem.
This is because gas density in clusters decreases as or faster 
than $\sim r^{-2}$ at the outer parts, so that the radiative cooling time 
in the $r\simgt 0.2r_{180}$ regions are much longer than the Hubble time.
Indeed, the results of numerical simulations for cluster formation 
show that the shapes of declining temperature profiles are 
essentially independent of the presence of radiative cooling and/or 
supernova feedback (e.g., \citealt{lok02,mot04}).  
For this reason, when we compare our model results with observations,
we pay attention to density and temperature profiles only in
cooling regions with $r\simlt 0.2r_{180}$ (typically $\sim0.5-1$ Mpc) 
beyond which the effect of radiative cooling is not serious
and thus heating mechanisms are not required.

The remainder of this paper is organized as follows:
In \S2 we evaluate the total heating rate resulting from the DF of galaxies
using the formula given by \citet{ost99}
for gravitational drag force in a gaseous medium.
While \citet{mil86} and EKK04 approximated galaxy motions as being
highly supersonic,  we allow for both subsonic and supersonic 
motions of galaxies.
In \S3 we calculate equilibrium
density and temperature profiles of galaxy clusters by assuming that 
the DF-induced heating is deposited at the locations of galaxies.
Effects of distributed heating of DF and the radial mass variation 
of galaxies are discussed in \S4.
In \S5 we analyze local thermal stability of the gas while in 
\S6 we present the results of numerical hydrodynamic simulations
that investigate the evolution of initially out of equilibrium configurations. 
Finally, in \S7 we conclude this work and discuss other potential
consequences of DF of galaxies in clusters.

\section{Total Heating Rate}

Galaxies orbiting around the center of a galaxy cluster gravitationally 
induce wakes in the ICM.  The gravitational 
interaction between the galaxies and wakes causes the former to lose
their orbital kinetic energy and converts it into thermal energy of
the ICM via either compressional heating (including shock) or
turbulent dissipation of the gas kinetic energy in the wakes.  
Although the notion that DF of galaxies can 
generate heat in the hot ICM has been well recognized and 
widely studied, most of previous work on this subject assumed 
galaxies in highly supersonic motions, estimated only
total heating rate, and compared it with total radiative loss rate 
in clusters
(\citealt{rud71,rep80,mil86}; EKK04).
However, the motions of galaxies are only slightly
supersonic, with an average Mach number of about 1.5 
(e.g., \citealt{sar88,fal04}), and even become subsonic
at the outer parts of clusters. 
In this paper we consider both subsonic and supersonic galaxy motions 
and adopt the general formula of \citet{ost99} for the DF force
in a gaseous medium.

Following \citet{ost99}, we consider a galaxy of mass $M_g$ moving at 
velocity ${\bf v}$ through a uniform medium with density $\rho$ and sound 
speed $c_s$. The dynamical-friction force that the galaxy experiences is 
given by
\begin{equation} \label{Fdf}
\FDF = -\frac{4\pi\rho(GM_g)^2I}{v^3} {\bf v},
\end{equation}
where the efficiency factor $I$ is defined by
\begin{eqnarray} \label{I}
I \equiv 
\left\{ \begin{array}{ll}
\frac{1}{2} \ln\left(\frac{1+\cal{M}}{1-\cal{M}}\right) - \cal{M},&
{\cal M}<1 \\
\frac{1}{2} \ln\left(1-{\cal M}^{-2}\right) + 
\ln\left({vt}/{r_{\rm min}}\right),&
{\cal M}>1,
\end{array} \right.
\end{eqnarray}
with ${\cal M}\equiv v/c_s$ being the Mach number of the galaxy motion and 
$r_{\rm min}$ the effective size of a galaxy \citep{ost99}.
Note that for ${\cal M}\gg1$, equations (\ref{Fdf}) and (\ref{I}) with
$vt=r_{\rm max}$ corresponding to the maximum system size,
become identical to the \citet{cha43} 
formula for the drag force due to collisionless particles.
Although equation (\ref{Fdf}) is valid for a perturber 
moving on rectilinear trajectory through a uniform-density medium, 
numerical simulations show that it is a good approximation even 
in a radially-stratified, spherical system \citep{san01}.

We assume that the orbital velocities of $N_g$ galaxies that contribute to 
heating of the gas are described by the Maxwellian distribution,
\begin{equation} \label{Maxwell}
f(v) = \frac{4\pi N_g}{(2\pi\sigma_r^2)^{3/2}}v^2e^{-v^2/(2\sigma_r^2)},
\end{equation}
with a one-dimensional velocity dispersion $\sigma_r$. 
The total heating rate due to the DF is then given by
\begin{equation} \label{Edot}
\langle\dot{E}\rangle = 
N_g \langle -\FDF \cdot {\bf v} \rangle =
\frac{4\pi\rho(GM_g)^2N_g}{c_s}
\left\langle\frac{I}{\cal M}\right\rangle,
\end{equation}
where the angular brackets denote an average over equation (\ref{Maxwell}).

\begin{figure}
\epsscale{.80}
\plotone{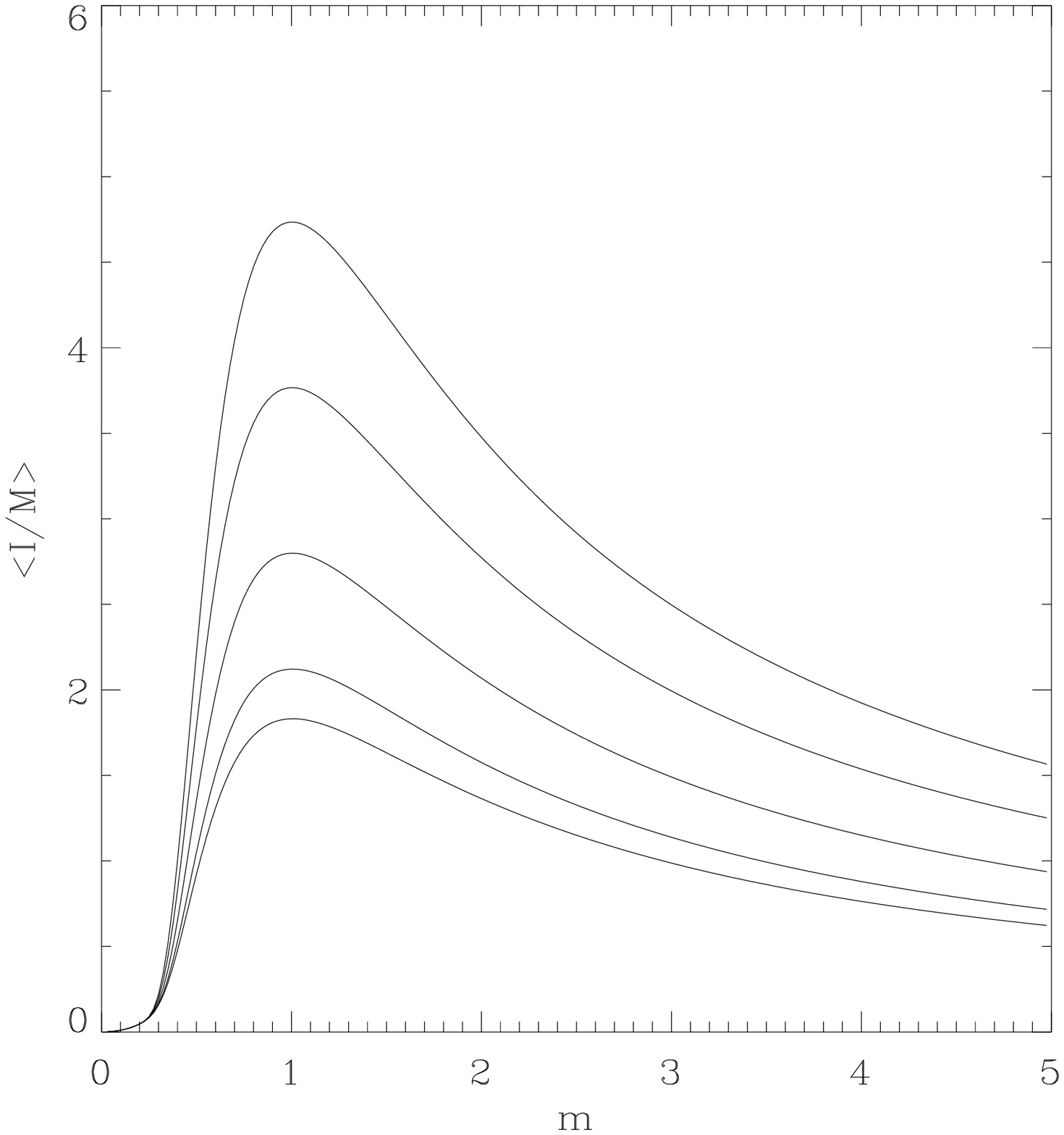}
\caption{Plots of $\langle I/{\cal M}\rangle$ as functions of 
$m\equiv \sigma_r/c_s$. 
Various curves corresponds to $\ln{(vt/\rmin)}=4, 4.6, 6, 8, 10$ 
from bottom to top.
\label{I_over_M}}
\end{figure}

Using equations (\ref{I}) and (\ref{Maxwell}), one can evaluate
$\langle I/{\cal M}\rangle$ numerically. Figure \ref{I_over_M} plots
$\langle I/{\cal M}\rangle$
as functions of $m\equiv \sigma_r/c_s$ 
for some values of $\ln{(vt/\rmin)}$. 
For supersonic cases, the density perturbations in the wake are 
highly asymmetric and the regions influenced by the perturber shrink
as $m$ increases.  This results in a smaller heating rate for larger $m$.
In this case, one can show $\langle I/{\cal M}\rangle \approx
(2/\pi)^{1/2}m^{-1}\exp(-0.5/m^2)\ln{(vt/\rmin)}$ for $m\gg1$.
For a subsonically moving perturber, on the other hand,
the perturbed density in the front and back sides of the perturber 
becomes more or less symmetric, producing a smaller heating rate with 
decreasing $m$.
In the limit of vanishingly small $m$, 
$\langle I/{\cal M}\rangle \rightarrow m^2$.
In general, $\langle I/{\cal M}\rangle$ peaks at $m=1$, as Figure
\ref{I_over_M} indicates.

Taking $vt=1$ Mpc from a typical size of clusters and $\rmin=10$ kpc
as the galaxy size, $\ln{(vt/\rmin)}\approx 4.6$.  
Since $m$ usually varies from 0.8 to 3,
$\langle I/{\cal M}\rangle\approx 2$.
Therefore, we have the total heating rate 
\begin{equation} \label{num}
\langle\dot{E}\rangle \approx4\times 10^{44}\;{\rm ergs\;s^{-1}}
\left(\frac{N_g} {500}\right)
\left(\frac{M_g} {10^{11}\; \Msun}\right)^2
\left(\frac{n_e} {0.01\pcc}\right)
\left(\frac{T} {5 \;\rm keV}\right)^{-1/2}
\left(\frac{\langle I/\mathcal M \rangle}{2}\right),
\end{equation}
where $n_e$ and $T$ denote the electron number density and
gas temperature, respectively, and we adopt the solar abundance.
Notice that the rate of gas heating given in equation (\ref{num}) 
is similar to typical X-ray luminosity of rich clusters
\citep[e.g.,][]{sar88,ros02}, 
implying that there is a sufficient amount of
energy available in the orbital motions of galaxies.
This is essentially what led EKK04 to conclude that
the dynamical-friction coupling between cluster galaxies and
gas can provide thermal energy enough to compensate for radiative loss.
However, it is still questionable 
whether DF of galaxies heats the gas in a right manner.
That is, are the observed density and temperature profiles of the ICM 
a direct consequence of energy balance between heating by DF and 
radiative cooling?  Is the intracluster gas heated by DF 
thermally stable?
In what follows, we shall study models of clusters with DF-induced 
heating in detail by assuming hydrostatic equilibrium and thermal 
energy balance.

\section{Model With Localized Heating}

DF of galaxies is mediated by gravity which is a long-range force, so
that heating of gas due to a single galaxy is likely to be well
distributed throughout its wake.  The functional form of heat distribution
is unknown and its derivation may require numerical simulations,
which are beyond the scope of the present work.
In this section, we instead make a simplifying assumption that
heat is deposited {\it locally} at the galaxy position.
The effects of heat distribution will be discussed in \S4.

\subsection{Local Heating Function}

Consider a spherically symmetric distribution of galaxies,
with the number density given by $n_g(r)$ at radius $r$. 
The local heating rate per unit volume due to DF is given by
\begin{equation} \label{edotr}
\edotr = 
\frac{4\pi\rho(GM_g)^2\IM}{c_s} n_g(r),
\end{equation}
where we assume equal galaxy mass. 
To find $n_g(r)$, we assume that the orbits of galaxies are isotropic and 
isothermal. 
The usual Jeans equation (e.g., eq.\ [4-55] of \citealt{bin87}) then reads
\begin{equation} \label{Jeans}
\sigma_r^2 \frac{d\ln n_g}{dr} = 
-\frac{d\Phi}{dr},
\end{equation}
where $\Phi$ is the gravitational potential.

Under the NFW distribution of dark matter
\begin{equation} \label {rho_dm}
\rho_{\rm NFW} = \frac{M_0/2\pi}{r(r+r_s)^2},
\end{equation}
with a scale radius $r_s$ and a characteristic mass $M_0$, 
the gravitational acceleration is given by 
\begin{equation} \label{dphi_dr}
-\frac{d\Phi}{dr} = 
-\frac{2GM_0}{r_s^2}
\left[\frac{\ln(1+x)}{x^2} - \frac{1}{x(1+x)}\right],
\end{equation}
where $x\equiv r/r_s$ is the dimensionless radius \citep{nav97,kly01}.
Combining equations (\ref{Jeans}) and (\ref{dphi_dr}) together, 
one can easily find 
\begin{equation} \label{ng}
n_g(x) = n_g(0)g(x) = N_g \frac{g(x)}{r_s^3\int g(x')d^3x'},
\end{equation}
where 
\begin{equation}\label{g}
g(x) \equiv \left[\frac{(1+x)^{1/x}}{e}\right]^\eta,
\end{equation}
with the dimensionless parameter $\eta$ defined by
\begin{equation} \label{eta}
\eta \equiv \frac{2GM_0}{r_s\sigma_r^2}.
\end{equation}
Note that $\eta$ measures the ratio of the gravitational potential 
energy of a galaxy in a cluster to that galaxy's 
kinetic energy at $r=r_s$.

\begin{figure}
\epsscale{.80}
\plotone{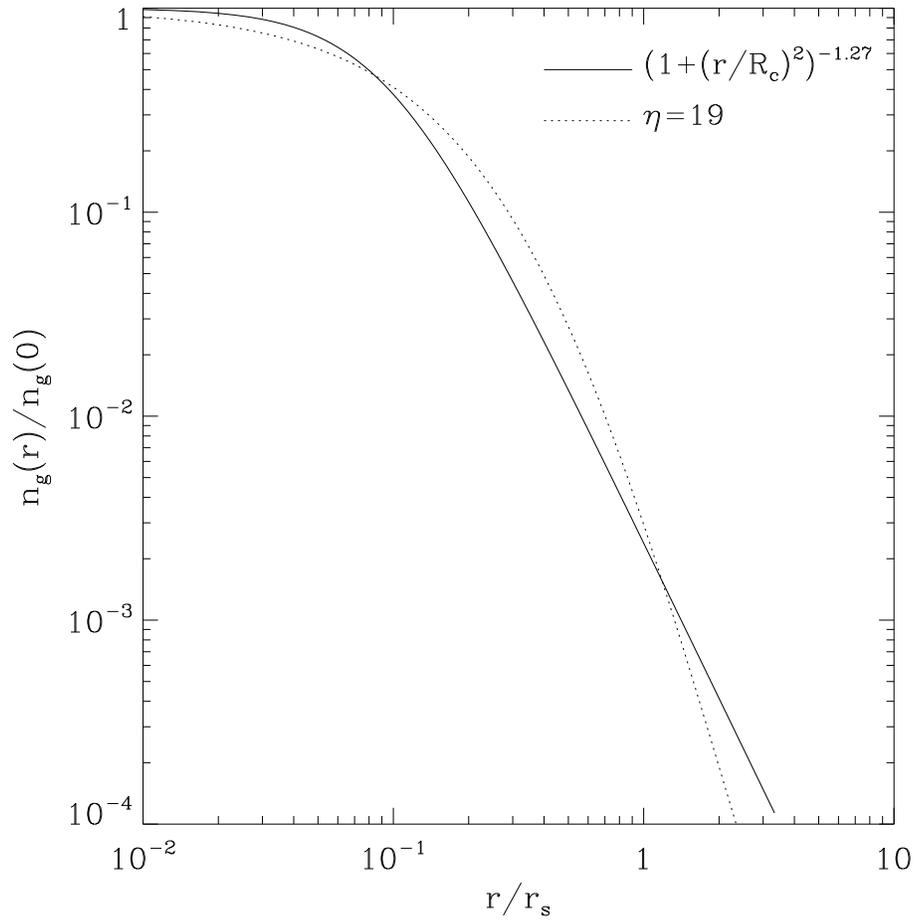}
\caption{Comparison of $g(r)$ with $\eta=19$ 
({\it dotted}) with the observed distribution ({\it solid})
of galaxies in A1795. The core radius is taken to be $R_c=43$ pc 
\citep{gir98}.
\label{den_comp}}
\end{figure}

While the functional form of $g(x)$ looks strange, it behaves quite well
near $x\sim0$ and gives density profiles similar to
the isothermal $\beta$-model of gas distributions 
\citep{mak98} or to the King model of the observed 
galaxy distributions \citep[e.g.,][]{gir98}.
For example, \citet{gir98} found that the best-fit distribution
of galaxies in A1795 is $n_g(r)=n_g(0)[1+(r/R_c)^2]^{-1.27}$ with
a core radius $R_c=43$ pc, which is plotted as a solid line in 
Figure \ref{den_comp}. On the other hand, X-ray and optical 
observations indicate $M_0=6.6\times10^{14}\Msun$, $r_s=460$ kpc, and
$\sigma_r\approx800\kms$ for A1795 \citep{gir98,ett02}, corresponding to
$\eta\approx 19$. Figure \ref{den_comp} plots as a dotted line
the $g(x)$ curve with $\eta=19$,
which is in good agreement with the observed galaxy distribution.

We take $2r_s$ as the outer boundary of a cluster.
Inserting equation (\ref{ng}) into equation (\ref{edotr}),
one obtains
\begin{eqnarray} \label{edot}
\edotx &=& 
\frac{4\pi\rho(GM_g)^2N_g}{c_sr_s^3\int g(x')d^3x'}
\left\langle\frac{I}{\cal M}\right\rangle g(x)
\nonumber \\
&=& 1.7\times 10^{-25}
r_{s,460}^{-3}
N_{g,500}M_{g,11}^2 n_e \TkeV^{-1/2}\IM g(x) \ergcms,
\end{eqnarray}
where $r_{s,460} = r_s/(460$ kpc), $N_{g,500}=N_g/500$,
$M_{g,11} = M_g/(10^{11}\Msun)$, and 
$\TkeV$ is the temperature of the gas in units of keV.
Equation (\ref{edot}) is our desired equation for the volume heating rate
due to the DF of galaxies.
Note that $\IM\approx2$ at $m\sim1$ and depends on temperature.

\subsection{Equilibrium Model}

We look for cluster density and temperature distributions in which
the hot gas satisfies both hydrostatic equilibrium and energy balance
between radiative cooling and DF-induced heating.
For thermal bremsstrahlung, the rate of energy loss per unit volume is 
given by
\begin{equation}\label{j}
j = 7.2\times 10^{-24} n_e^2\TkeV^{1/2}\ergcms,
\end{equation}
\citep{ryb79,zak03}.
From equations (\ref{edot}) and (\ref{j}),
the condition of thermal energy balance yields
\begin{equation} \label{neT}
n_e\TkeV = 2.4 \times10^{-2} r_{s,460}^{-3}N_{g,500}M_{g,11}^2 \IM g(x).
\end{equation}
Neglecting the weak temperature dependence of $\IM$ and
assuming that $M_g$ does not vary with radius, equation (\ref{neT})
states that the gas pressure in equilibrium should trace the distribution
of galaxies.

Hydrostatic equilibrium of the gas requires
\begin{equation} \label{HSE}
\frac{c_0^2}{n_e}\frac{d(n_e\TkeV)}{dr} = 
- \frac{d\Phi}{dr},
\end{equation}
where $c_0 = 390\kms$ is an {\it isothermal} sound speed at $\TkeV=1$. 
Substituting equations (\ref{dphi_dr}) and (\ref{neT}) into 
equation (\ref{HSE}) and using equation (\ref{g}), we obtain
\begin{equation} \label{diff}
\frac{d}{dx}\ln\IM = -\eta 
\left(\frac{\sigma_r^2}{c_0^2\TkeV} - 1 \right)
\left[\frac{\ln(1+x)}{x^2} - \frac{1}{x(1+x)}\right].
\end{equation}
Since $\IM$ depends on $\TkeV$, equation (\ref{diff}) can 
be integrated as an initial value problem to yield $\TkeV(x)$.
The equilibrium profile for $n_e(x)$ can then be found from equation 
(\ref{neT}).
Since $n_e(x)$ depends rather sensitively on less-well known 
quantities  such and $N_g$ and $M_g^2$ and 
is easily influenced by the mass profile of galaxies, we 
concentrate on $\TkeV(x)$ that is independent of the properties of
galaxies other than $\sigma_r$.

Equation (\ref{diff}) has a trivial isothermal solution 
\begin{equation} \label{Tsol}
\Tiso
= 6.5 
\left(\frac{\sigma_{r}}{10^3\kms}\right)^2\;\;{\rm keV},
\end{equation}
independent of the radius.  
The corresponding density profile is
\begin{equation} \label{nsol}
n_e = 7.4\times10^{-3} r_{s,460}^{-3}
\sigma_{r,3}^{-2} N_{g,500}M_{g,11}^2 
g(x),
\end{equation}
where $\sigma_{r,3}=\sigma_{r}/(10^3 \kms)$, indicating
that the gas density follows the galaxy distribution exactly.
Considering uncertainties in the values of $r_{s,460}$, $\sigma_{v,1000}$, 
$N_{g,500}$, and $M_{g,11}$,\footnote{For example, 
the average mass of galaxies would increase up to 
$\sim 5\times 10^{11}\Msun$ if the contribution from dark halos
is taken into account \citep[e.g.,][]{zen03}.}
equation (\ref{nsol}) gives $n_e\sim 0.01-0.2\pcc$ at the cluster 
centers, which is not much different from observed values.
Note that the Mach number corresponding to $\Tiso$ is 
$m=\gamma^{-1/2}$.
For later purposes, we define the transonic temperature
\begin{equation}\label{Tsonic}
\Tsonic \equiv \frac{1}{\gamma}\Tiso,
\end{equation}
corresponding to unity Mach number of galaxy motions.

The isothermal solution (\ref{Tsol}) is consistent with what
has long been known as the scaling relation between the gas temperature 
and the velocity dispersion of galaxies for many clusters
(e.g., \citealt{mus84,edg91,lub93,wu99}, see also \citealt{sar88,ros02} 
for review).  It explains why the values of the
$\beta$ $(\equiv \mu m_p \sigma_r^2/kT)$ parameter in the $\beta$ models
for gas distributions are close to unity \citep{lub93,bah94,gir96b,gir98}.
Physically, the scaling relation implies that the plasma and the 
galaxies are well relaxed under the common gravitational potential.  
It is uncertain whether the observed scaling law results truly from 
the DF coupling between the ICM and galaxies, but 
equation (\ref{Tsol}) suggests that the latter certainly
makes the former tighter.

\begin{figure}
\epsscale{1.00}
\plotone{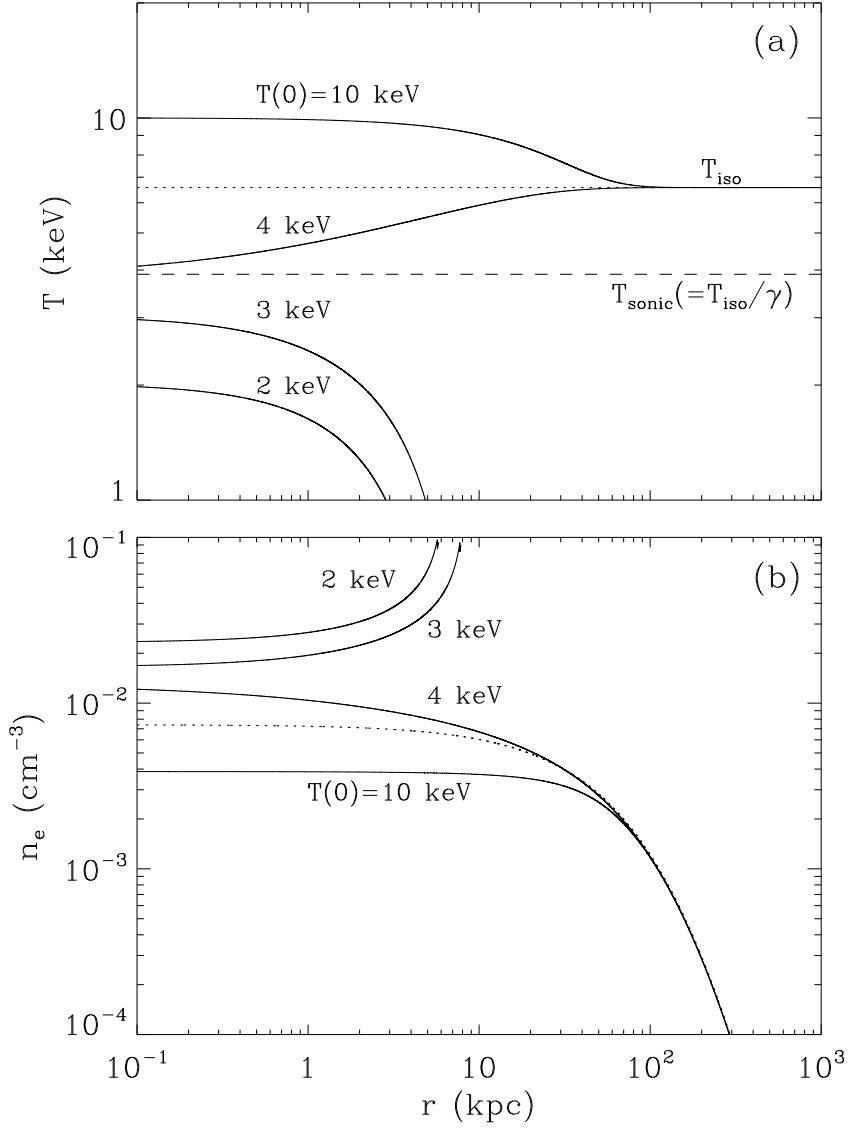}
\caption{
({\it a}) Equilibrium temperature and 
({\it b}) electron density profiles of clusters in
which DF-induced heating is spatially localized.
The chosen parameters are $\eta=19$, $r_s=460$ kpc, $N_g=500$,
$M_g=10^{11} \Msun$ and $\sigma_r=10^3\kms$.
Dotted and dashed lines in ({\it a}) respectively indicate
the characteristic isothermal solution $\Tiso=6.5$ keV
and the transonic temperature $\Tsonic\equiv\Tiso/\gamma=3.9$ keV 
at which the Mach number 
of galaxy motions is unity. 
The density distribution corresponding to the isothermal solution
is drawn as a dotted line in ({\it b}).
Note that temperature increases with radius only 
when $\Tsonic<T(0)<\Tiso$. 
\label{local}}
\end{figure}

Although gas in the regions surrounding cooling cores of clusters has nearly
constant temperatures close to the virial values, recent X-ray observations 
exhibit positive radial gradients of gas temperatures 
at the central $\sim300$ kpc regions \citep{pet01,pet03},
while showing declining temperature distributions at far outer regions
\citep{mar98,deg02,pif05,vik05}.
Thus, the isothermal solution cannot describe the observed temperature
distributions for the whole range of radii. 
We are particularly interested in the regions within and adjacent to 
the cooling radius in which gas density is high enough to experience
significant radiative cooling.
To check whether the general solutions of equation (\ref{diff}) produce 
temperature and density distributions similar to the observed in
such regions, we solve it numerically.
We adopt
$\eta=19$, $r_s=460$ kpc, $N_g=500$, $M_g=10^{11} \Msun$, and 
$\sigma_r=10^3\kms$ corresponding to 
$\Tiso= 6.5$ keV. We choose a value for the central temperature $T(0)$
and then integrate equation (\ref{diff}) from $r=0$ to $2r_s$.
The resulting temperature and density profiles for a few 
selected values of $T(0)$ are plotted in Figure \ref{local}. 

The spatial behavior of the solutions depends critically on $T(0)$.
When $T(0)>\Tsonic$, galaxy motions are subsonic and 
$d\IM/dm>0$ from Figure \ref{I_over_M}.
If $T(0)>\Tiso$ (or $m<\gamma^{-1/2}$), equation (\ref{diff}) gives
$d\IM/dr= -(m/2T)(d\IM/dm)(dT/dr)>0$ for $T(0)>\Tiso$,
implying $dT/dr<0$.  
As temperature monotonically decreases with radius, $m$ increases until it 
reaches the isothermal value $\gamma^{-1/2}$ where $T=\Tiso$. 
One can similarly show that $dT/dr>0$ for $\Tiso>T(0)>\Tsonic$,
and thus temperature slowly increases toward $\Tiso$.
As long as $T(0)>\Tsonic$, the solutions asymptote to 
the isothermal ones (eqs.\ [\ref{Tsol}] and [\ref{nsol}]) as 
the radius increases.   On the other hand, $dT/dr<0$ when
$T(0)<\Tsonic$ (or $m>1$). 
Since $m$ increases with decreasing $T$ and since
$d\IM/dr$  does not change its sign for $m>1$, 
$dT/dr<0$ is satisfied for all radii.  The decrease of
temperature is much faster than that of $g(x)$, resulting in
a unrealistic distribution of electron number density that
increases with radius.

As Figure \ref{local} shows, the local heat balance with DF-induced heating
yields rising temperature profiles, if and only if $\Tsonic<T<\Tiso$.
While this is a tantalizing result considering the
central depression of temperatures seen in cooling-flow clusters,
the required range of temperatures is very narrow.
If we identify $\Tiso$ with the virial temperature, 
the central temperatures in our equilibrium models must be 
larger than $\Tsonic=0.6\Tiso$ (for $\gamma=5/3$), 
which is about twice larger than the observed values of 
$T(0)\sim(0.3-0.4)\Tiso$ (e.g., \citealt{pet01,pet03,all01}).
By having too stringent upper and lower limits of temperatures,
therefore, heating by DF of galaxies alone is 
unlikely to explain observed temperature and density distributions 
of galaxy clusters. 
We note however that the tight range of temperatures for
the rising temperature profiles may be caused by the basic 
assumptions of this section, namely that 
DF-induced heating of the ICM is all localized 
at the positions of galaxies, and that all galaxies have equal mass.
We relax these assumptions in the next section.

\section{Effects of Distributed Heating}

In the presence of viscosity and/or turbulent magnetic fields in the ICM,
the kinetic energy lost by a galaxy via DF will eventually be 
converted into heat, rather than bulk motions distributed  onto the  
gravitational wakes of the galaxy.
\citet{dei96} used a quasi-linear fluctuation
theory to derive the spatial structure (in Fourier space) 
of mechanical heating in a turbulent medium 
self-consistently driven by DF of many galaxies.
To our knowledge, there is no published work that
study spatial heat distribution (in real space) caused by DF.
If ICM heating by DF of galaxies occurs through 
turbulent dissipation of gas kinetic energy, finding the functional form of 
the heat distribution would not be viable
unless the characteristics of ICM turbulence and related
processes are prescribed \citep{dei90}.
Instead of attempting to derive a realistic heat distribution,
in this work we simply adopt a Gaussian function to parametrize 
the spatial extent to which heat is distributed\footnote{We also
tried with logarithmic heat distribution functions
used in EKK04 and found that results are similar to those
with Gaussian functions presented in this section.}. 
Our aim in this section is to examine the effects of heat 
distribution on equilibrium structures in comparison
with the localized heating models.
Since masses of galaxies inside a cluster are likely to vary with the 
radius within the cluster,
we also allow for spatially varying galaxy masses.

Assuming that heat distribution follows a Gaussian profile
centered at the location of the galaxy,
we write the heating rate per unit volume as 
\begin{equation} \label{dist_edotr}
\edotr =
\frac{4\pi\rho G^2\IM}{c_s} h(r),
\end{equation}
where 
\begin{equation} \label{conv}
h(r) \equiv
\frac{1}{\pi^{1/2} l_h} \int n_g(r')M_g(r')^2 e^{-(r-r')^2/l_h^2} dr',
\end{equation}
is a convolution of $n_gM_g^2$ with the Gaussian
function with scale length $l_h$.
In writing equations (\ref{dist_edotr}) and (\ref{conv}), we assume that
density and temperature of the gas do not change
significantly inside the wake of a galaxy, which is acceptable within the 
framework of equation (\ref{Fdf}).
The condition of hydrostatic equilibrium and thermal energy
balance is then reduced to
\begin{equation} \label{diff2}
\frac{d}{dx}\ln\IM = -\eta 
\frac{\sigma_r^2}{c_0^2\TkeV} 
\left[\frac{\ln(1+x)}{x^2} - \frac{1}{x(1+x)}\right] 
- \frac{d\ln h(x)}{dx}.
\end{equation}
It can be easily verified that equation (\ref{diff2}) recovers 
equation (\ref{diff}) in the limit of $l_h\rightarrow 0$.

\begin{figure}
\epsscale{1.00}
\plotone{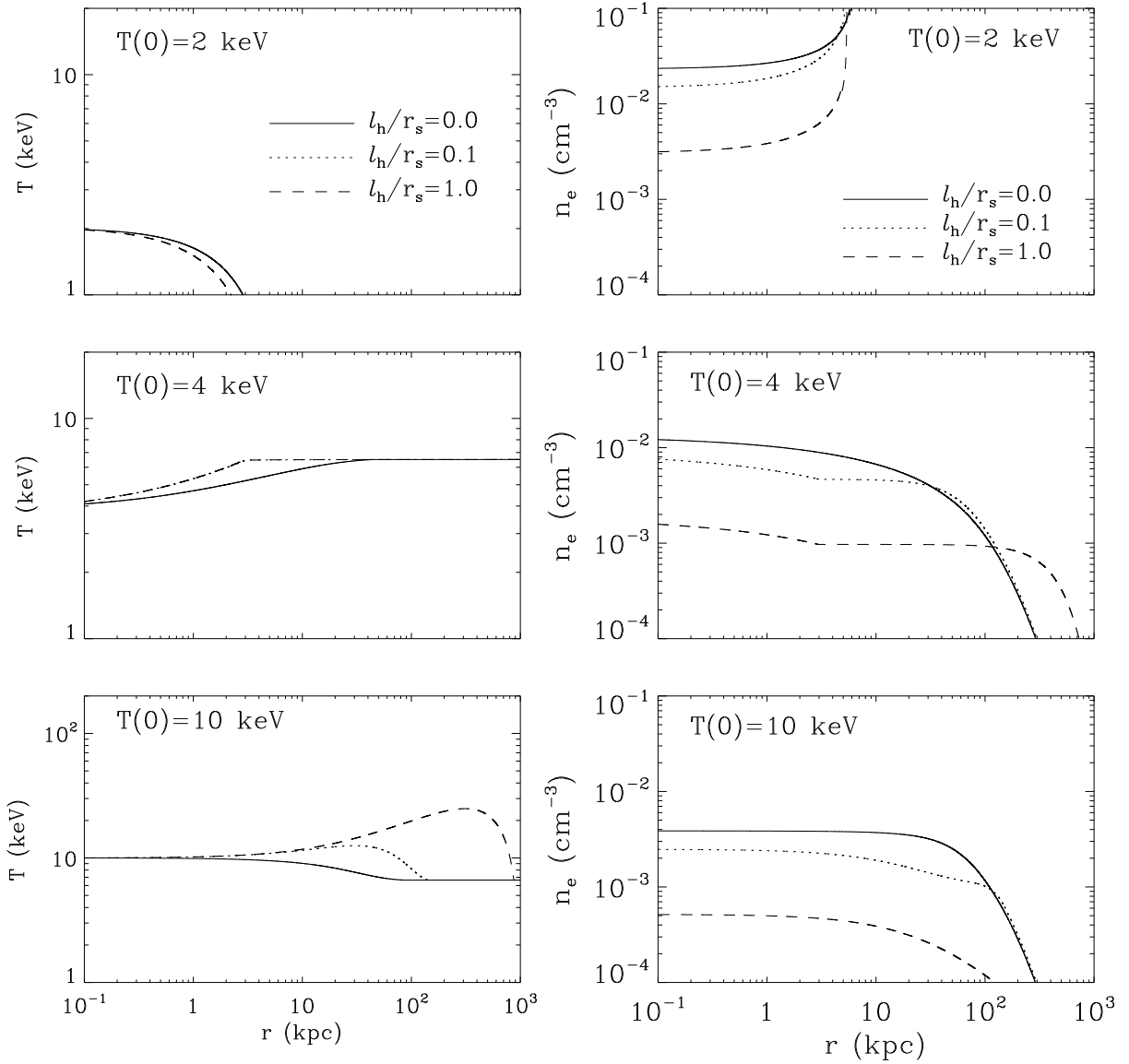}
\caption{
Equilibrium (left) temperature and (right) density profiles of cluster 
models with distributed heating,
which is modeled by a Gaussian function with scale length $l_h$.
The chosen parameters are the same as in Figure \ref{local}.
When $T(0)=2$, 4 keV, temperature profiles with $l_h/r_s=0.1$ and 1.0
are identical to each other.
See text for discussion.
\label{disth}}
\end{figure}

Figure \ref{disth} plots the solutions of equation (\ref{diff2})
for the case of constant $M_g$.
More widely distributed heating is equivalent to localized heating with
a flatter distribution of galaxy number density,
causing the $|d\ln h/dx|$ term in equation (\ref{diff2}) to be smaller.
Consequently, an equilibrium cluster with a larger value of $l_h$ tends to
make $\IM$ decreases faster with increasing $r$.
This implies that for $T(0)<\Tsonic$ ($T(0)>\Tsonic$), 
the temperature in models
with non-zero $l_h$ decreases (increases) more rapidly than in the 
$l_h=0$ counterpart, 
although the difference between models with $l_h/r_s=0.1$ and 1.0 
is negligible for small $r$.
Note that even when $T(0)>\Tiso$, for which localized-heating models
have declining temperature profiles, equilibrium temperature in
distributed-heating models increases
in the regions with $r<l_h$ and eventually converges to $\Tiso$ at 
$r\simgt(2-3)l_h$.
We see that distributed heating does not change the condition
$\Tsonic<T<\Tiso$ for the existence of rising temperature distributions.
The distributed-heating models do not work
better than the localized-heating model in terms of producing 
temperature distributions similar to observations;
it in fact aggravates the situation by making all of the
gas in clusters nearly isothermal except only in the central $\sim 1$ kpc 
regions.

Another factor that may affect the equilibrium structure is the radial mass 
variation of cluster galaxies.
Although galaxies located near the central parts of clusters tend to have
a large fraction of luminous matter, (e.g., \citealt{biv02,mer03}), 
the strong tidal stripping of dark-matter halos is likely to
cause the total (luminous + dark) masses of individual galaxies to be 
smaller toward the cluster center (\citealt{zen03}; EKK04; \citealt{nag05}).
Motivated by this consideration, we adopt several algebraic forms of
$M_g(r)$ that either monotonically decrease or increase 
by about a factor of 20 or less from the cluster center to $r=2r_s$,
and then solve equation (\ref{diff2}).  
The resulting temperature profiles are almost 
identical to those of the constant-mass cases presented in Figure \ref{disth}.  
Because the distribution of galaxy masses appears 
logarithmically in equation (\ref{diff2}), 
it does not considerably affect temperature structures,
although it dramatically changes electron number-density profiles.

\section{Thermal Stability}

\begin{figure}
\epsscale{1.00}
\plotone{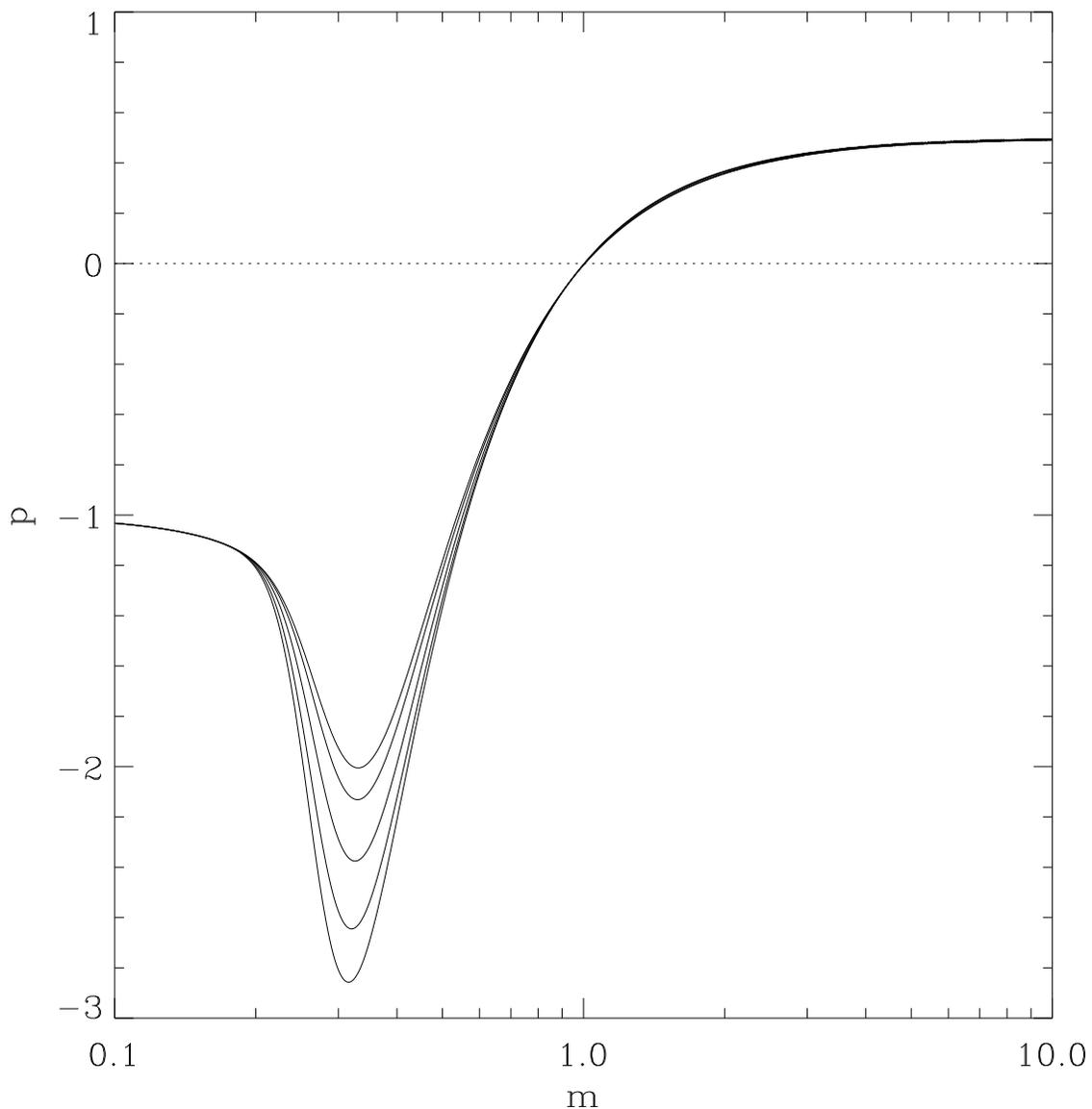}
\caption{
Logarithmic slope $p\equiv d\ln\IM/d\ln T
=-0.5 d\ln\IM/d\ln m$ as a function of Mach number $m$
for $\ln{(vt/\rmin)}=4, 4.6, 6, 8, 10$ from top to bottom.  
The curves asymptote to $-1$ for $m\ll1$ and to $0.5$ for $m\gg1$.
With positive values of $p$, 
DF-induced heating by supersonically moving galaxies with $m>1$ 
is unable to suppress thermal instability.
\label{slope}}
\end{figure}

We now discuss thermal stability of a system in which radiative cooling
is locally balanced by DF-mediated heating.  For
spatially localized heating,
the net loss function $\rho\calL$ is given by
\begin{equation} \label{calL}
\rho\calL = j-\edot = \alpha \rho^2T^{1/2} - \beta \rho T^{-1/2+p},
\end{equation}
where the power index $p$ accounts for the local temperature 
dependence of $\IM$ in a piecewise manner, and the positive constants 
$\alpha$ and $\beta$ contain all information 
on the cluster properties other than gas density and temperature.
Figure \ref{slope} plots $p=d\ln\IM/d\ln T$ as a function of $m$, which
gives $p\approx 0$ at around the transonic temperature 
($m\approx 1$), while $p\approx 0.3-0.5$ for supersonic temperatures ($m>1$). 
The curves asymptote to $-1$ for $m\ll1$ and to $0.5$ for $m\gg1$,
as the asymptotic formulae given in \S2 suggest. 
Note that $p$ is insensitive to the choice of $\ln(vt/r_{\rm min})$ 
as long as $m>0.8$.

Thermal stability of a system can be easily checked by using
the generalized Field criterion which states that 
the system is thermally unstable to isobaric perturbations if
\begin{equation}\label{gfield}
\left.\frac{\partial (\calL/T)}{\partial T}\right|_P
= \frac{1}{\rho T}
\left[
\left.\frac{\partial (\rho\calL)}{\partial T}\right|_\rho
   -  \frac{\rho}{T}
\left.\frac{\partial(\rho\calL)}{\partial \rho}\right|_T
\right] = 
- p \frac{\alpha \rho}{T^{3/2}} < 0,
\end{equation}
where the second equality assumes $\rho\calL=0$ corresponding to
thermal equilibrium \citep{fie65,bal86}.
Since $p>0$ for $m>1$ (and $p<0$ for $m<1$),
this implies that the ICM is thermally unstable (stable) if heating is
caused preferentially by galaxies moving at a supersonic (subsonic) 
speed. 
Dense inner parts of galaxy clusters, where the cooling time is less than 
the Hubble time, are filled with low-temperature gas such that galaxy motions 
are readily supersonic, implying that DF-induced heating is 
unable to quench thermal instability in those cooling cores.
Whether thermal instability has important dynamical consequences on
cluster evolution can be judged by considering its growth time which 
amounts to
\begin{equation}\label{tgrow}
t_{\rm grow} = -\frac{\gamma}{\gamma-1} \frac{P}{\rho T^2}
\left.\frac{\partial(\calL/T)}{\partial T}\right|_P^{-1}
=0.96\; p^{-1}\; {\rm Gyr}\; 
\left(\frac{n_e}{0.05 \pcc}\right)^{-1} 
\left(\frac{k_BT}{2\; {\rm keV}}\right)^{1/2},
\end{equation}
(cf.\ \citealt{kim03a}).  
Even if cooling cores are thermally unstable, therefore,
the virulence of thermal instability may or may not manifest 
itself during the 
lifetime of clusters depending on the value of $p$.
For example, for a cluster with $\sigma_r=10^3\kms$ and $T=2$ keV
at the central regions, $m=1.4$ and 
$p=0.23$ from Figure \ref{slope}, giving $t_{\rm grow}=4.2$ Gyr.
This is almost comparable to the ages of clusters since the
last major merger event ($\sim7$ Gyr for massive clusters; \citealt{kit96}),
suggesting that thermally instability may be dynamically unimportant
for practical purposes. 

The inability of DF-induced heating to suppress thermal 
instability seems not to be a fatal problem as long as
the following two conditions are met:
i) cooling cores are in thermal equilibrium, a basic premise of
the stability analysis;
and ii) galaxies near the central parts move at near-transonic speeds 
($m\simlt2$), so as to have a sufficiently small value of $p$.  
While the second condition appears to be easily satisfied in clusters, 
the results of the preceding section suggests that 
the first condition may not be so if DF-induced heating is to
balance radiative cooling for the {\it whole} range of radii.
Since the cooling time outside the cooling radius is longer than
the Hubble time,  one may argue that material beyond the cooling radius
does not need any heating source and that it is sufficient for DF of galaxies 
to supply heat to gas only in cooling cores. 
We address this possibility in the next section.

As a final remark of this section, we compare our results with previous work.  
Formally, equation (\ref{gfield}) implies that 
gas subject to DF-induced heating has a critical temperature above (below) 
which it becomes 
thermally stable (unstable) and that this critical temperature 
corresponds exactly to unity Mach number of galaxy motions.
This result is quite different from those of
previous work which made some approximations on the heating function.  
For instance, \citet{mil86} argued that gas with drag heating
is always thermally unstable and rapidly turns into a multi-phase medium,
a consequence of his neglecting the temperature dependence of the 
heating function (corresponding to $p=1/2$ in eq.\ [\ref{calL}]).
Although \citet{bre89} found a similar critical gas temperature, they
considered only the supersonic regime and used an approximate heating 
function from the shock jump conditions, which makes their
critical temperature smaller than ours by a factor of about three. 
Our result, based on the general formula of \citet{ost99}
for the drag force applicable for both supersonic and subsonic cases,
represents the situation better.

\section{Time Dependent Approach}

We have shown in the preceding sections that an equilibrium cluster subject 
to DF-mediated heating has a temperature profile that either decreases
(for $T>\Tiso$ or $T<\Tsonic$) or increases (for $\Tsonic<T<\Tiso$)
with radius.  While the radially increasing temperature
profile is attractive, the requirement that thermal energy is balanced
locally in the entire regions out to $2r_s$
causes the central gas temperature to be no less 
than 0.6 
times the virial temperature, about a factor two larger than observations.  
Since radiative cooling is important only in the dense central parts inside 
the cooling radius, however, it would be sufficient for DF of galaxies to 
provide heat only in such a cooling core instead of the entire regions.
In addition, thermal instability would not be an issue provided that gas 
is in thermal equilibrium and that galaxy motions are near transonic (\S5). 
Therefore, it may be possible for a cluster to start from an arbitrary 
non-equilibrium state and evolve slowly under both radiative cooling
and DF-mediated heating, ending up with an equilibrium core
in which thermal instability has a very long growth time.
The resulting temperature profile,
albeit consistent with observations, need not represent an equilibrium
for the whole range of radii.

\begin{figure}
\epsscale{.80}
\plotone{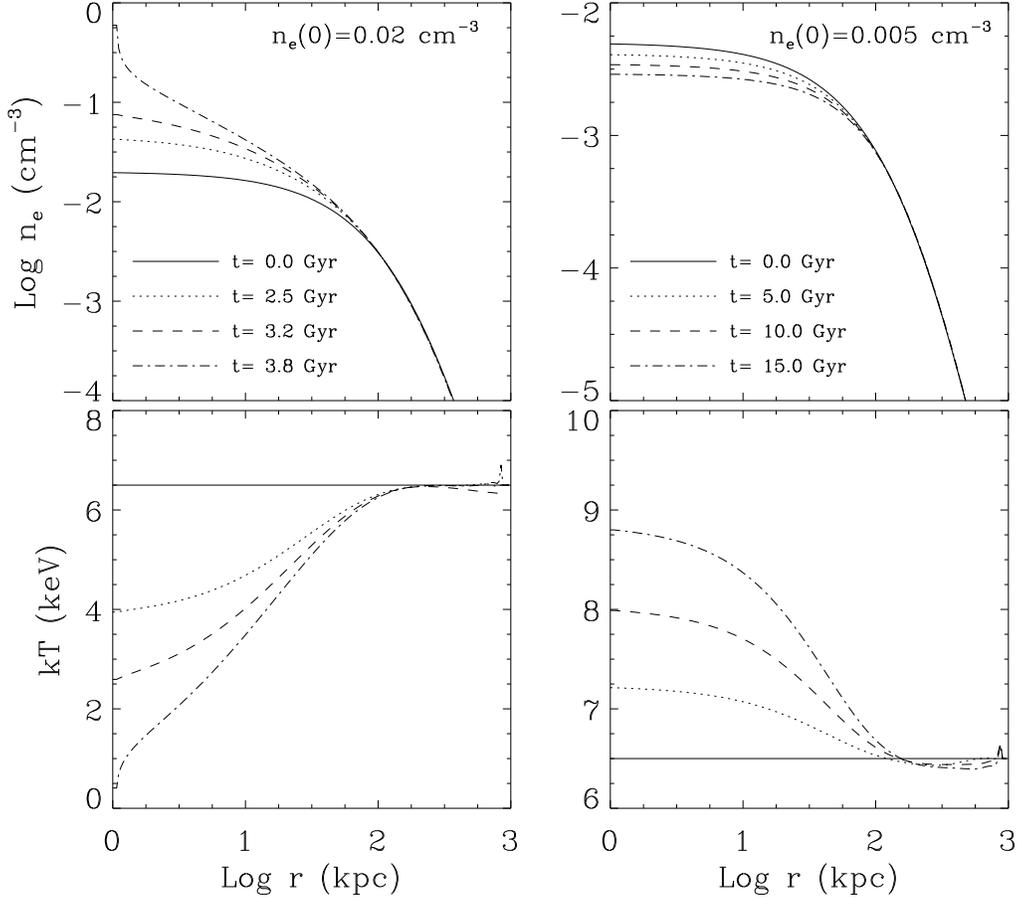}
\caption{
Evolution of electron number density ({\it top}) and temperature
({\it bottom}) for ({\it left}) cluster model A with $n_e(0)=0.02\pcc$ 
and ({\it right}) model B with $n_e(0)=0.005\pcc$.  The initial
temperature is 6.5 keV for both models.
Model A, which initially has a larger density than the equilibrium value,
inescapably experiences a cooling catastrophe, 
while model B with lower initial density steadily heats up.
\label{evol}}
\end{figure}

To test this idea, we have run a number of one-dimensional 
hydrodynamic simulations in spherical polar coordinates.
We set up a logarithmically spaced radial grid, with 400 zones,
from 1 kpc to 1 Mpc, and explicitly implement the net cooling-heating 
function described by equation (\ref{calL}). 
We also include the effect of gravitational potential due to 
{\it passive} dark matter using equation (\ref{dphi_dr}) with 
$M_0=6.6\times 10^{14}\Msun$ and $r_s=460$ kpc; 
the DF-coupling between gas and the dark matter and back reaction on 
the latter component are not taken into account. 
As initial conditions, we consider spherically symmetric clusters and 
take $n_e=n_e(0)g(x)\pcc$ with
the central electron density $n_e(0)= 0.002$, 0.005, 0.01, 0.02 cm$^{-3}$, 
and fix $T=6.5$ keV independent of the radius for all models. 
We adopt $r_s=460$ kpc, $N_g=500$, and $M_g=10^{11} \Msun$, and 
consider both localized heating and distributed heating with
$l_h/r_s=$0.1 and 1.
Note that the corresponding equilibrium density profile has
$n_e(0)= 0.0074$ cm$^{-3}$ when heating is localized.
We adopt the outflow boundary conditions for scalar variables 
(i.e., density, energy, etc.) that assign the same values in the ghost 
zones as in the corresponding active zones.  For the radial velocity,
we allow it to vary as a linear function of radius at the inner
boundary, while fixing it to be zero at the outer boundary 
(e.g., \citealt{kim03a}).  
Using the ZEUS hydrodynamic code \citep{sto92}, we 
solve the basic hydrodynamic equations  
and follow the nonlinear evolution of each model cluster.  
The resulting radial profiles of electron number density and temperature
of model A with $n_e(0)=0.02\pcc$ and model B with $n_e(0)=0.005\pcc$,
both of which assume spatially localized heating,
are shown in Figure \ref{evol} for a few time epochs.

Model A, which initially has larger density than the equilibrium
value everywhere,  immediately develops radial mass inflows.  
As time evolves, the temperature drops and the density increases. 
Compared to pure cooling cases which fairly well maintain near
isobaric conditions (e.g., \citealt{kim03a}),
DF-induced heating is found to cause thermal pressure to increase
with density as $P\propto\rho^{0.3}$.
Since $j\propto T^{-2.4}$ and $\edot \propto T^{-1.9+p}$
with $p>0$ for supersonic galaxy motions, however,
the system is thermally unstable and 
cooling occurs at a much faster rate as temperature decreases.
Model A experiences a cooling catastrophe in less than 4 Gyr.
On the other hand, model B with initial overheating becomes hotter and 
less dense, which in turn tends to increase the ratio of heating
rate to the cooling rate for $p\simgt -0.5$.  
As the gas temperature increases, the Mach number of 
galaxy motions become smaller, reducing the value of $p$.
The cooling rate will therefore eventually exceed the heating rate and 
the cluster evolution will be reversed.  In the case of model B, this 
turnaround will
occur at $t=70$ Gyr when $T\sim18$ keV (or $m\sim0.5$) is reached.

The evolution of models with different density and different length scales 
for heat distribution is qualitatively similar to those of models A and B, 
namely that clusters catastrophically evolve toward vanishing central 
temperature if cooling dominates initially, while clusters with initial 
excessive heating heat up steadily.
This suggests that DF-induced heating does not naturally lead 
non-equilibrium clusters to thermally stable, equilibrium cores.
As shown in \S3 and \S4,  DF of galaxies does not explain the observed
temperature distributions of clusters if the condition of 
thermal energy balance is imposed for all radii.  
Although it is enough for DF-induced heating 
to balance radiative cooling only in cooling cores, 
such cooling cores do not naturally form from non-equilibrium states.
Unless the properties and galaxies and ICM are fine tuned, 
small departures from an equilibrium state rapidly evolve into an extreme 
configuration.  Therefore, 
we conclude that DF-induced heating {\it alone} is not likely to account 
for the absence of cold gas in the centers of galaxy clusters.

\begin{figure}
\epsscale{1.00}
\plotone{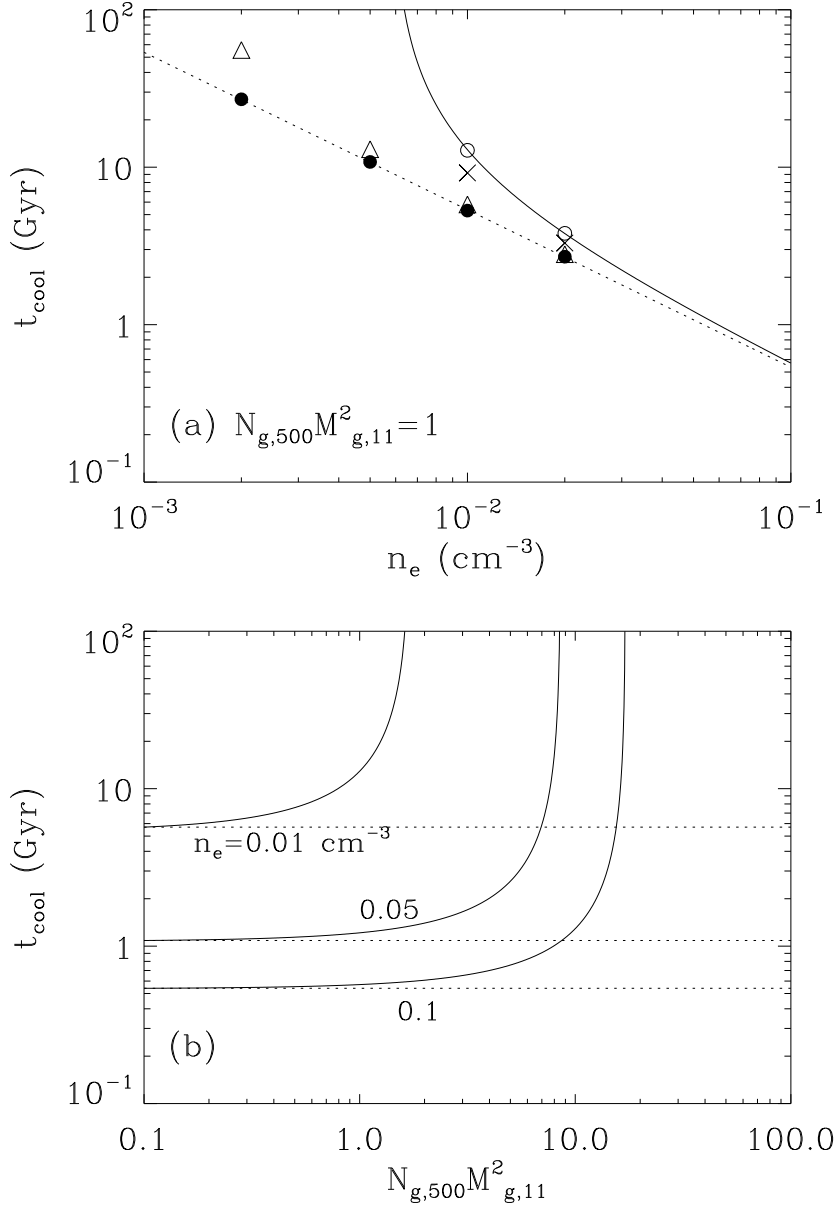}
\caption{Cooling time scale at the cluster center ({\it a}) as a function 
of $n_e$ for fixed $N_{g,500} M_{g,11}^2=1$ and ({\it b}) as a function
of $N_g M_g^2$ for $n_e=0.1, 0.05, 0.01\pcc$.
Solid line is for the case
when spatially localized, DF-induced heating 
is considered, while dotted line corresponds to the case
without any heating.
Various symbols in ({\it a}) represent
simulation outcomes, with solid circle, triangle, cross, and open
circle referring to no heating, distributed heating with
$l_h/r_s=1.0$ and 0.1, and localized heating cases, respectively.
\label{tscale}}
\end{figure}

Although heating by DF from galaxies does not appear to provide a complete 
solution to the cooling
flow problem, we see that the DF-induced heating can still offset a
considerable amount of radiative cooling.   
The isobaric cooling time is given by
\begin{equation}\label{tcool}
t_{\rm cool} = \frac{\gamma}{\gamma-1}
\left(\frac{P}{\rho\calL}\right) 
= \frac{0.96}{1-\mathcal{R}/(n_e\TkeV)} 
\left(\frac{n_e}{0.05\pcc}\right)^{-1} 
\left(\frac{kT}{2\;{\rm keV}}\right)^{1/2}
\;{\rm Gyr},
\end{equation}
where $\mathcal{R} \equiv 0.024 r_{s,460}^{-3} N_{g,500} M_{g,11}^2\IM$
represents the contribution of the DF-induced heating.
Figure \ref{tscale} plots as solid lines the cooling time as a function of 
$n_e$ for $N_{g,500} M_{g,11}^2=1$ or 
as a function of $N_{g,500} M_{g,11}^2$ with varying density.  
The cases without any heating are compared as dotted lines.  
The time epochs when clusters experience cooling catastrophe in the numerical 
simulations are marked in Figure \ref{tscale}$a$ as solid circle, triangle, 
cross, and open circle for no heating, the distributed heating 
with $l_h/r=1.0$ and 0.1, and 
the localized heating cases, respectively.  
The numerical results are in good agreement with the theoretical prediction.  
Models with distributed heating tend to 
have longer cooling time. 
When $N_{g,500} M_{g,11}^2=1$ and $n_e(0)=0.05\pcc$, Figure \ref{tscale}$a$ 
shows that DF-induced heating lengthens the cooling time by about 13\% 
compared to the pure cooling case.
However, as Figure \ref{tscale}$b$ indicates, the offset of cooling 
by DF-induced heating is increasingly larger as $N_{g,500}M_{g,11}^2$ 
increases.  The increment of the cooling time could be
as large as 140\% when $n_e(0)\sim0.03-0.1\pcc$ and 
$N_{g,500}M_{g,11}\sim3-7$, suggesting that thermal energy supplied
by DF of galaxies is by no means non-negligible. 

\section{Conclusions and Discussion}

Friction of galaxy motions via the gravitational interaction with their own
gravitationally induced wake in the ICM has often been invoked as an
efficient heating mechanism of the gas
(e.g., \citealt{mil86,jus90,dei96}; EKK04).  
This idea of DF-mediated heating of the ICM is quite attractive
because there is sufficient energy available in galaxy motions and 
the mechanism is self-regulating; it operates effectively only when the 
temperature of gas is in a certain range, which happens to be the typical
gas temperatures in the cooling cores of galaxy clusters.
In this paper, we take one step further from \citet{mil86} and EKK04
to calculate equilibrium density and temperature profiles of the hot gas
heated by DF of galaxies.
%Our primary objective is to investigate whether DF-induced heating alone
%is responsible for the absence of cooling flows in the cores of 
%rich clusters.  
Instead of restricting to cases where galaxy motions are all supersonic,
we use the general formula derived by \citet{ost99} for the drag force 
that takes allowance for both supersonic and subsonic galaxy motions
in a gaseous medium.
We show that the total heating rate due to the DF of galaxies in a 
typical rich cluster is comparable to its total X-ray luminosity,
confirming the results of \citet{mil86} and EKK04 (see \S2). 

Next, we derive the local heating function (eq.\ [\ref{edot}])
assuming that the orbits of galaxies are isotropic and isothermal
under the NFW distribution of dark matter
and that the kinetic energy lost by a galaxy is 
deposited into heat at the location of the galaxy.
The condition that the gas is in hydrostatic equilibrium and 
maintains energy balance requires the temperature profile to be 
one of the following three kinds:
isothermal, with $T=\Tiso$; a decreasing profile with radius 
when $T<\Tsonic$ or $T>\Tiso$;
an increasing profile when $\Tsonic<T<\Tiso$, 
where $\Tiso$ and $\Tsonic$ denote the temperatures
corresponding to unity Mach number of galaxy motions with respect 
to the isothermal and adiabatic sound speeds, respectively (eqs.\ 
[\ref{Tsol}] and [\ref{Tsonic}]).
The isothermal solution 
%that stems presumably from the assumption
%of isotropic and isothermal galaxy orbits 
is interesting because it quite
well describes the observed scaling relationship among clusters properties.
Although the rising temperature solution is attractive since
clusters usually show a temperatures drop at the central regions,
it strictly requires the central temperature to be no lower than 0.6 times 
the virial temperature, which is roughly twice smaller than the 
observed values (\S3).
We also consider cases in which DF-induced heating is distributed in space
according to a Gaussian form, and/or the masses of cluster galaxies vary
over radius, and find that 
the stringent limit of $T>\Tsonic=0.6\Tiso$ 
for rising temperature distributions remains unchanged (\S4).

Using the local heating function we have derived, we analyze thermal 
stability of the gas subject to radiative cooling and 
DF-mediated heating (\S5).  
When galaxy motions are subsonic,
the heating rate that deceases steeply with temperature
suppresses thermal instability completely.
On the other hand, supersonic galaxy motions for which
the heating rate is relatively insensitive to temperature
are unable to stop the growth of thermal instability.
The growth time in the presence of DF-induced heating is at least 
twice that in the pure cooling case, and becomes progressively 
longer as the Mach number of galaxies decreases. 
When galaxies move at slightly supersonic velocities
with Mach number less than 2, which is very
likely at cluster centers, the growth time becomes comparable to the ages
of clusters. 
This implies that thermal instability, even if operates,  may have
insignificant dynamical consequences on cluster evolution.

Noting that regions outside the cooling radius need not be in strict 
thermal equilibrium,  we look for a possibility using numerical hydrodynamic 
simulations that clusters evolve from an arbitrary 
non-equilibrium state and form current cooling cores in which
DF-induced heating balances radiative cooling (\S6).
We find that clusters that were initially dominated by cooling
unavoidably develop radial mass inflows and decreases
their central temperatures in a runaway fashion,
whereas clusters with initial overheating slowly heat up and
result in radially decreasing temperature profiles.
Equilibrium solutions therefore do not appear to form an 
attracting set for galaxy and gas configurations,
suggesting that when DF from galaxies is the sole 
heating source, it is extremely difficult to obtain equilibrium cores
by smoothly evolving non-equilibrium clusters (even if in some cases
the cooling catastrophe can be deferred so as to occur on a longer timescale).

Putting together all the results of this paper
we conclude that DF of galaxies 
{\it alone}, albeit an interesting heating mechanism with a lot of
available energy, 
cannot be the lone heating agency to balance radiative cooling
in rich galaxy clusters.  
We nonetheless note that the heating due DF of galaxies
could considerably lengthen the cooling time, depending on the value 
of $N_g M_g^2$,
and thus should not be neglected in energetics of
galaxy clusters.

One of the key assumptions made in this paper is that all 
the kinetic energy lost by galaxies via DF is transferred to the thermal 
energy of the surrounding gas.
The rationale behind this assumption is that the superposition of 
the wakes produces turbulence in the ICM and the kinetic energy 
injected into the ICM turbulence at saturation cascades down along 
the Kolmogorov-like energy spectrum, 
and eventually transforms into heat through viscous dissipation 
at small scales.
Another possibility is that a large fraction of the kinetic energy of 
galaxies is used to merely enhance the level of the turbulence
instead of being converted into heat. 
This may occur
when the turbulence is not fully developed yet.

The characteristics of the ICM turbulence is not well known,
yet observations and numerical simulations suggest an average
velocity dispersion $\vturb\sim 200-400\kms$ on scales of
$\lambda\sim5-20$ kpc \citep{ric01,car02,chu04,sch04,fal04}.
The associated turbulent kinetic energy is 
$(1/2)M_{\rm gas}\vturb^2 \sim 9\times 10^{61}$ erg for 
the total mass $M_{\rm gas}\sim 10^{14}\Msun$ of the ICM in a rich cluster.
Let us assume that this energy is supplied solely by DF of galaxies at a rate 
given by equation (\ref{num}). 
If the level of turbulence keeps increasing without dissipation, 
it would take about 7 Gyr for the DF of galaxies to feed the
ICM turbulence to the observed level,
which is almost comparable to the lifetime of the cluster.
Conversely, if the turbulence is fully developed and in a steady state 
such that the energy injection rate by DF is equal to the rate
of dissipation, it would have
$\vturb = (2\lambda\dot{E}/M_{\rm gas})^{1/3}\sim 60 \kms$
for $\lambda\sim 20$ kpc, which is too small to explain the
observations.
All of these imply that the contribution from the DF of galaxies to
the ICM turbulence is not considerable (see also \citealt{san99b}). 
If the dissipation of the (fully developed) turbulence is to provide 
enough heat to balance radiative cooling, as suggested by \citet{chu04}
and \citet{fuj04}, the energy injection should not be entirely due to
the DF of galaxies; it requires other energy sources, including jets from
active galactic nuclei and mergers with smaller groups or clusters.

The assertion that the conversion of galaxy kinetic energy
into thermal energy of the gas may not be so effective is supported by the 
results of \citet{fal04} who studied using numerical simulations 
cluster formation in $\Lambda$CDM cosmology, 
with the DF of galaxies included explicitly.
They found that 
(1) the motions of galaxies at the present epoch are slightly
supersonic, with an average Mach number of ${\cal M}\approx 1.4$;
(2) gas within the virial radius has a velocity dispersion
$\vturb\sim(0.3-0.5)c_s$ resulting probably from infall motions
of galaxies and small groups, with small contribution from the DF of
galaxies; and (3) the clusters still suffer from the cooling catastrophe.
Although higher-resolution, more-controlled simulations are required
to make a definitive statement, 
the last point of their results
suggests that the system is thermally unstable
or heating by DF may be quite inefficient 
--- for, as we have seen in the previous section, systems evolved from 
arbitrary initial conditions do not generally tend to the equilibrium 
solutions, which 
do not therefore form an attracting set; and for some initial parameters 
the delay in the cooling catastrophe that the DF mechanism leads to 
is not significant.

We finally note that several issues of potential importance were not 
investigated in this paper. One of these relates to the galaxy velocity 
anisotropy. It is easy to show that, for isothermal distribution with 
constant anisotropy, it is possible to obtain equilibrium solutions with 
smaller central gas temperatures, more in line with observations.
However these tend to have unrealistic galaxy number density distributions. 
The  equilibrium and stability of gas configurations in more realistic 
anisotropic models have not been investigated.
We have  adopted the general formula of \citet{ost99} for 
the drag force on a galaxy moving at an arbitrary speed.
This is an improvement over previous
studies (e.g., \citealt{mil86}; EKK04) that considered
only supersonic galaxy motions.
While its explicit dependence on the Mach number enabled us to explore
temperature profiles of equilibrium clusters,
the formula still assumes a galaxy moving in a {\it straight-line} orbit 
through a {\it uniform} gaseous medium. 
Clearly, the ICM is radially stratified and galaxies follow
curved rather than rectilinear trajectories. 
In a collisionless system, \citet{jus04} recently showed that 
the density inhomogeneity reduces the Coulomb logarithm of the 
Chandrasekhar formula by limiting the the maximum impact parameter 
to the local scale length of density variation 
(see also \citealt{has03,spi03}),
and induces an additional drag force in the lateral direction of
the galaxy motion (see also \citealt{bin77}).
They further showed that the reduction of the Coulomb logarithm
makes the orbital energy loss 15\% less effective, inhibiting the 
circularization of the orbit, while the additional tangential drag
force has a negligible effect on the orbital evolution.
Similar effects are expected to occur in a gaseous medium, yet
their consequences are not known quantitatively.
A related important question is how heat (or turbulent) energy is 
distributed in the wakes that form in the turbulent, inhomogeneous,
magnetized ICM.  A real assessment of the effects of DF to dynamical 
evolution of
galaxy clusters requires answers to the above questions, which will 
direct our future research.

\acknowledgments

We are pleased to thank P. Goldreich, M.\ G.\ Lee, 
C.\ McKee, R.\ Narayan, 
and E.\ Ostriker for many helpful discussions and communications.
We also thank an anonymous referee for useful comments
and suggestions.
W.-T.\ Kim was supported in part by Korea Science and Engineering 
Foundation (KOSEF) grant R01-2004-000-10490-0 at the SNU 
and in part by NSF grant AST 0307433 at the CfA.
M.\ Kamionkowski was supported by NASA NNG05GF69G at Caltech.

\clearpage

%fig1 

%fig2 

%fig3 

%fig4 

%fig5 

%fig6 

%fig7 

\end{document}